\documentclass[conference]{IEEEtran}
\IEEEoverridecommandlockouts
\usepackage{cite}
\usepackage{amsmath,amssymb,amsfonts}
\usepackage{algorithmic}
\usepackage{graphicx}
\usepackage{subfigure}
\usepackage{textcomp}
\usepackage{xcolor}
\usepackage{stfloats}
\usepackage{array}
\usepackage{bm}

\usepackage{gensymb}

\begin{document}
\title{An SBR Based Ray Tracing Channel Modeling Method for THz and Massive MIMO Communications}
\DeclareRobustCommand*{\IEEEauthorrefmark}[1]{%
	\raisebox{0pt}[0pt][0pt]{\textsuperscript{\footnotesize\ensuremath{#1}}}}
\author{\IEEEauthorblockN{Yuanzhe Wang\IEEEauthorrefmark{1}, Hao Cao\IEEEauthorrefmark{1}, Yifan Jin\IEEEauthorrefmark{1}, Zizhe Zhou\IEEEauthorrefmark{1}, Yinghua Wang\IEEEauthorrefmark{2}, Jialing Huang\IEEEauthorrefmark{1}, \\Yuxiao Li\IEEEauthorrefmark{1}, Jie Huang\IEEEauthorrefmark{1,2*} and Cheng-Xiang Wang\IEEEauthorrefmark{1,2*}}
	\IEEEauthorblockA{$^1$ National Mobile Communication Research Labratory, School of Information Science and Engineering,\\ Southeast University, Nanjing 210096, China \\ $^2$ Purple Mountain Labratories, Nanjing 211111, China \\$^*$Corresponding Authors: Jie Huang, Cheng-Xiang Wang\\
		Email: vincewong202612@163.com, \{213201162, 213200781, 213193095\}@seu.edu.cn, \\wangyinghua@pmlabs.com.cn, \{jlhuang, yuxli, j\_huang, chxwang\}@seu.edu.cn}}
		
\maketitle
\begin{abstract}
	Terahertz (THz) communication and the application of massive multiple-input multiple-output (MIMO) technology have been proved significant for the sixth generation (6G) communication systems, and have gained global interests. In this paper, we employ the shooting and bouncing ray (SBR) method integrated with acceleration technology to model THz and massive MIMO channel. 
	The results of ray tracing (RT) simulation in this paper, i.e., angle of departure (AoD), angle of arrival (AoA), and power delay profile (PDP) under the frequency band supported by the commercial RT software Wireless Insite (WI) are in agreement with those produced by WI. 
	Based on the Kirchhoff scattering effect on material surfaces and atmospheric absorption loss showing at THz frequency band, the modified propagation models of Fresnel reflection coefficients and free-space attenuation are consistent with the measured results. For massive MIMO, the channel capacity and the stochastic power distribution are analyzed. The results indicate the applicability of SBR method for building deterministic models of THz and massive MIMO channels with extensive functions and acceptable accuracy.
\end{abstract}
\section{Introduction}
The incoming 6G technology characterizes wider ranges of application scenarios, with transmission rates and communication quality rising significantly \cite{1}--[4]. Thus, an accurate channels model incorporating new technologies like THz communication and massive MIMO is indispensable to meet the requirements in 6G design and deployment.

In \cite{3}, the RT based deterministic modeling was introduced as a perfect candidate for wireless channel prediction. The two most commonly adopted approaches for RT are the image method (IM) and the SBR method. 
%The IM model tracks effective rays with higher accuracy while consuming tremendous amounts of time. 
The IM method can trace the ray with high efficiency while being time-consuming.  
By contrast, the SBR method can trace multiple rays simultaneous with width first search algorithm, therefore featuring lower computational complexity and better extensions \cite{4}. These advantages make it more versatile and efficient in complex simulations. 

In \cite{4}--[9], THz and massive MIMO were proved to be prospective solutions to increase transmission rate. THz features ultra-high transmission rate with larger channel bandwidth available. In \cite{5}, the Kirchhoff scattering theory was introduced to revise the diffusion of rays in THz band.  In \cite{6}, attenuation and dispersion of atmosphere during the propagation of rays in THz channels were proved to be significant.
Massive MIMO improved channel capacity and spectral efficiency, while it features the characteristics of near-field spherical wavefront when the distance between the transmitting and receiving antennas or a cluster are probably within the Rayleigh distance \cite{7}. Under such circumstances, the simulation and calculation methods of MIMO channels need to be modified accordingly.

Recently, the most common approach to model high-frequency channels and massive MIMO channels is the geometry based stochastic model (GBSM), which is versatile but requires tremendous channel measurement, as well as being less accurate in specific scenarios than RT method \cite{8}--[11]. WI, the most popular commercial RT software recently, is also limited in the simulation of high frequency channels and it only supports the band between 50MHz to 100GHz. So in this paper the SBR method is adopted to perform THz and MIMO channel simulation, and the application of RT method in channel simulation can also be easily extended as a result of that.

This paper focuses on the SBR based THz and massive MIMO channel modeling, and it innovatively combines THz correction with the SBR simulation structure, as well as verifying several important properties of massive MIMO with the SBR method. The atmospheric absorption and Kirchhoff scattering effect at THz band are considered and MIMO arrays are equipped at both transmitter and receiver sides of the channels. The simulation system is set up based on MATLAB and is to be extended into an integrated RT module for 6G channels in the future.

The rest of the paper is organized as follows. Section II briefly reviews the SBR based channel modeling methods for THz band and massive MIMO channels. Section III describes the implement on MATLAB of the proposed channel model. In Section IV we present the simulation and analysis results. The conclusions are drawn in Section V.
\section{SBR BASED CHANNEL MODELING FOR THZ BAND}
\subsection{SBR method based forward RT and calculation}\label{2.1}
The SBR method is employed in the simulation, whose procedure consists of four steps, i.e., ray launching, ray tracing, ray reception, and ray calculation \cite{3}.

\subsubsection{Ray launching}\label{2.1.1} 
\begin{figure}[tb]
	\centering
	\includegraphics[width=0.3\textwidth]{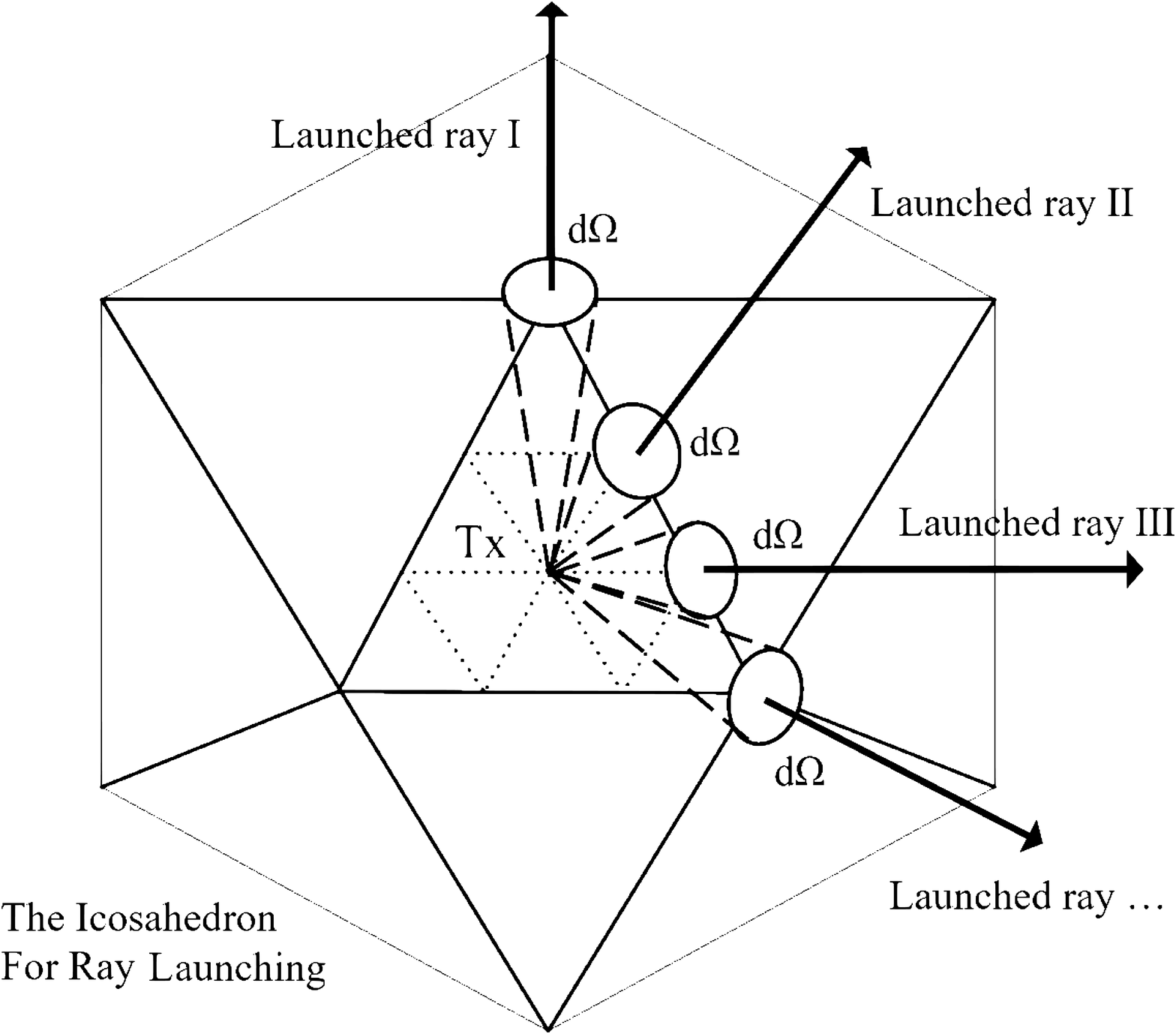}
	\caption{The icosahedron sampling method in ray launching.} \label{A}
\end{figure}

Previous works have demonstrated the effectiveness of icosahedron ray-sampling method, where an icosahedron is created to assist in the launch of rays. This method, illustrated in Fig.~\ref{A}, characterizes the uniformity of launched rays' distribution as well as provides significant topological information about different rays launched.

\subsubsection{Ray tracing}\label{2.1.2} 
\begin{figure}[tb]
	\centering
	\subfigure[]{
		\includegraphics[width=0.3\textwidth]{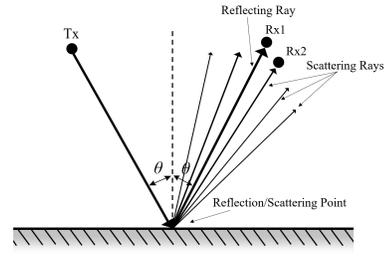}
	}
	\qquad
	\subfigure[]{
		\includegraphics[width=0.3\textwidth]{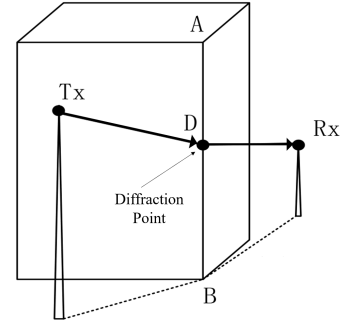}
	}
	\caption{Reflection and scattering in (a) and diffraction in (b).}

	\label{B}
\end{figure}

In this step, each ray is tracked at the given propagation mechanisms (the permitted times of reflection, diffraction, scattering, etc.) according to Snell's law and ER theory as it propagates through geometric scenarios, and the different propagation mechanisms are illustrated vividly in Fig.~\ref{B}. Simultaneously, the rays will experience various power losses during their propagation in the scenarios. At the same time, RT trees are established to store and arrange for the propagation of ray cones, and then pick out effective paths for the next step in \ref{2.1.3}.

\subsubsection{Ray reception}\label{2.1.3} 
\begin{figure}[tb]
	\centering
	\includegraphics[width=0.3\textwidth]{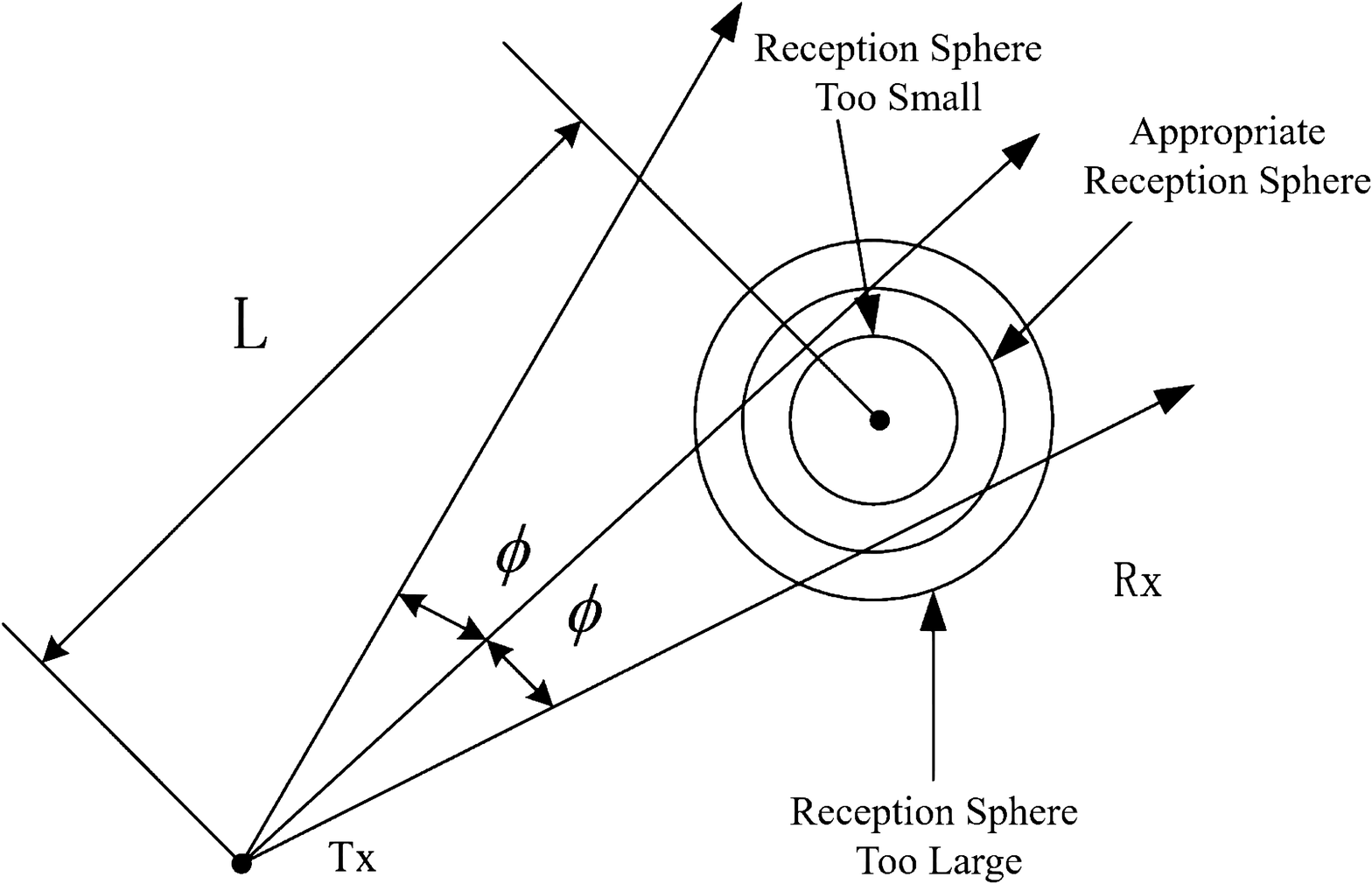}
	\caption{The icosahedron sampling method in ray launching.} \label{C}
\end{figure}

Effective paths selected in the last step in \ref{2.1.2} are tested to determine whether the wavefront intersects the receiving antenna. Reception spheres in Fig.~\ref{C} are used to model this process where the radius of spheres is determined by the distance between the signal transmitter and receiver, representing the cross section of ray cones. If a center ray intersects the receiver sphere, the cone it represents will contribute for the total electric field of the receiver.

\subsubsection{Ray calculation}\label{2.1.4}  With Maxwell equations in time-harmonic electromagnetic fields, the electric field $\vec{\bm{Er}}_{\theta,\varphi,i}$ of the $i$-th received ray is calculated according to the propagation record of it, which can be demonstrated by \cite{b}
\begin{align}
	{\vec{\bm{Er}}}_{\theta,\varphi,i}=&\left[\prod{\vec{\bm{R}}}_{\theta,\varphi,j}\right]\cdot\left[\prod{\vec{\bm{S}}}_{\theta,\varphi,k}\right]\notag \\
	&\cdot\left[\prod{A_{\theta,\varphi,l}\left(s^\prime,s\right)\cdot \vec{\bm{D}}}_{\theta,\varphi,l}\right]\cdot{\vec{\bm{A}}}_{d_i}\cdot{\vec{\bm{E}}}_{\theta,\varphi,i}
\end{align}
where $ {\vec{\bm{E}}}_{\theta,\varphi,i}$ is the initial electric field of the $i$-th received ray, ${\vec{\bm{A}}}_{d_i}$ is free space attenuation factor, $ {\vec{\bm{R}}}_{\theta,\varphi,j}$, ${\vec{\bm{S}}}_{\theta,\varphi,k}$, and $\vec{\bm{D}}_{\theta,\varphi,l}$ are the coefficients of the $j$-th reflection, $k$-th scattering, and  $l$-th diffraction, respectively. $A_{\theta,\varphi,l}\left(s^\prime,s\right)$ represents the attenuation function.

In massive MIMO channel, electric fields need be combined coherently with phase. The total electric field received at one certain receiving point $\vec{\bm{Er}}_{\theta,\varphi}$ is accumulated by ${\vec{\bm{Er}}}_{\theta,\varphi,i}$ according to the linear nature of Maxwell equations, which can be calculated as

\begin{equation}
	(Er)_{\theta,\varphi} = | \sum_{n=1}^{N_{p}}  \vec{\bm{Er}}_{\theta,\varphi ,n} | 
\end{equation}
where $N_p$ is the number of multipath components (MPCs).

The total power at the receiving point is
derived from the power density of the electromagnetic
 field, and it can be
\begin{equation}
	P=\frac{\lambda^2\beta}{8\pi\eta _0} (Er)_{\theta,\varphi}^2
\end{equation}
where $\lambda$ represents the wavelength, $\eta_0$ is the free space impedance, and $\beta$ is the quality factor.

The channel impulse response (CIR) can describe a channel, which is
\begin{equation}
	h\left(\tau\right)=\sum_{n=1}^{N_p}{A_n\delta\left(\tau-\tau_n\right)e^{-j\psi_n}}
\end{equation}
where $\delta(\cdot)$ is the Dirac function, $A_n$, $\tau_n$, and $\psi_n$ denote the amplitude, delay, and phase of the $n$-th path, respectively.

The multipath richness of a communication channel can be illustrated by the root mean square (RMS) delay spread, which can be described as
\begin{equation}
	\tau_{RMS}=\sqrt{\frac{\sum_{n=1}^{N_p}P_n\tau_n^2}{\sum_{n=1}^{N_p}P_n}-\left(\frac{\sum_{n=1}^{N_p}P_n\tau_n}{\sum_{n=1}^{N_p}P_n}\right)^2}
\end{equation}
where $\tau_n$ denotes the delay of the $n$-th path, $N_p$ is the total number of effective paths and $P_n$ is square of the amplitude of $n$-th path.
\subsection{Characteristics of wireless channels at THz band}
\subsubsection{The Kirchhoff scattering theory} Wavelength at THz frequencies is at the order of millimeters or below, thus the roughness of materials can not be neglected. In such environments, diffuse scattering can result in severe power loss in specular reflection directions. The Rayleigh roughness factor \cite{7} from Kirchhoff scattering theory is introduced as
\begin{equation}
	\rho=e^{-\frac{g}{2}}
\end{equation}
where $g$ is calculated by
\begin{equation}
	g=\left(\frac{4\pi\cdot\sigma\cdot cos{\theta_i}}{\lambda}\right)^2
\end{equation}

 $\sigma$ represents the roughness of materials, $\theta_i$ is the incident angle. Then the modified reflection coefficient can be demonstrated as
\begin{equation}
	r_{TE/TM}^\prime=\rho\cdot\ r_{TE/TM}
\end{equation}
where $r_{TE/TM}$ stands for the original Fresnel reflection coefficients under TE/M polarization, while $r_{TE/TM}^\prime$ for the modified ones at THz band.

\subsubsection{The attenuation and dispersion of atmosphere} At THz band, the molecular absorption effect of water vapor and oxygen is non-negligible. Thus, we adopt an algorithm to calculate atmospheric attenuation and dispersion. The specific atmospheric attenuation is given by \cite{8}
\begin{align}
	\gamma = &\gamma _o + \gamma _w = 0.1820 f  ( N^{\prime \prime}_{Oxygen}(f)+ \notag \\
	&N^{\prime \prime}_{Water\ Vapour}(f))\quad {\rm (dB/km)}
\end{align}
where $\gamma_o$ and $\gamma_w$ are the specific attenuation of dry air and water vapor, respectively. $f$ denotes the frequency. $N^{\prime \prime}\left(f\right)$  is the imaginary part of complex refractive index for oxygen and water vapor.

The specific atmospheric dispersion is given by
\begin{align}
	\varphi=&\varphi_o+\varphi _w=-1.2008f(N_{Oxygen}^\prime(f)+\notag \\
	&N_{WaterVapor}^\prime(f)) \quad {\rm (\degree /km)}
\end{align}
where $\varphi_o$ and $\varphi_w$ stand for specific phase dispersion of them. $N^\prime(f)$ is the real part of complex refractive index for oxygen and water vapor.
\subsection{The characteristics of wireless channels in massive MIMO scenarios}
At the working frequency band of RT method, the size of antenna can normally be ignored compared to the scenario. The antenna matrix of MIMO system constructed in uniform planar array (UPA) or uniform linear array (ULA) can be simplified as a series of points, denoted as \cite{7}
\begin{equation}
	\bm{T}_x = \left[T_1\quad T_2\quad \ldots \quad T_{N_t}\right]
\end{equation}
\begin{equation}
	\bm{R}_x = \left[R_1\quad R_2\quad \ldots \quad R_{N_r}\right]
\end{equation}
where $T_i(i=1,\ldots,N_t)$ and $R_j(j=1,\ldots,N_r)$ are the transmitting and receiving points in the MIMO arrays.

The channel matrix $H_{N_r\times N_t}$ is commonly employed to describe MIMO channels, and it can be
\begin{equation}
	{\bf H}_{N_r \times N_t} =\left(
	\begin{matrix}
			h_{11}     & h_{12} & \ldots & h_{1N_{t}}  \\
			h_{21}     & h_{22} & \ldots & h_{2N_{t}}  \\
			\vdots     & \vdots & \ddots & \vdots      \\
			h_{N_{r}1} & h_{N_{r}2} & \ldots & h_{N_r N_t} \\
		\end{matrix}
	\right)
\end{equation}
where $h_{qp}(q=1,\ldots,N_r,p=1,\ldots,N_t)$  denotes the CIR for transmit antenna $p$ and receive antenna $q$ and can be calculated as
\begin{equation}
	h_{qp}=\sqrt{\left(P_{qp}\right)}e^{i\theta_{qp}}
\end{equation}
where $P_{qp}$ is the received power of antenna $R_q$ from transmitting antenna $T_p$, and $\theta_{qp}$ denotes the phase component of $P_{qp}$. 

The channel capacity $C\left(SNR\right)$, which is applied to illustrate the maximum transmission rate of a MIMO channel. It is denoted as

\begin{equation}
	C(SNR)=log_2\left| {\bf I}_n + \frac{SNR}{N_t}{{\bf W}} \right|\quad \rm{bits/Hz}
\end{equation}
%\begin{equation}
	\begin{align}
		\bf{W} = &\left\{ \begin{array}{rl}
			Q\bf{H}\bf{H}^T  & N_t\ge N_r \\
			Q\bf{H}^T \bf{H} & N_t<N_r    \\
		\end{array}\right.\\
		%\quad
		&Q=\sum_{q=1}^{N_r}\sum_{p=1}^{N_t}||h_{qp}||^2
\end{align}
	%Q=\sum_{r=1}^{N_r}\sum_{t=1}^{N_t}||h_{rt}||^2
%\end{equation}
%\begin{equation}
%	Q=\sum_{r=1}^{N_r}\sum_{t=1}^{N_t}||h_{rt}||^2
%\end{equation}
where $SNR$ denotes the average signal-to-noise ratio, ${\bf H}^T$ stands for the Hermitian transpose of ${\bf H}$ and ${\bf I}_n$ is the identity matrix.

\begin{figure}[tb]
	\centering
	\includegraphics[width=0.5\textwidth]{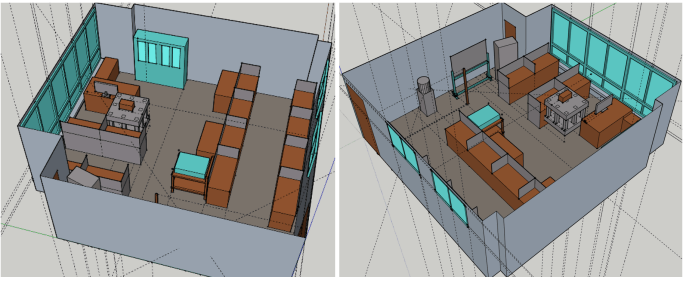}
	\caption{The simulated office scenario.} \label{1}
\end{figure}

\begin{figure}[bt]
	\centering
	\includegraphics[width=0.5\textwidth]{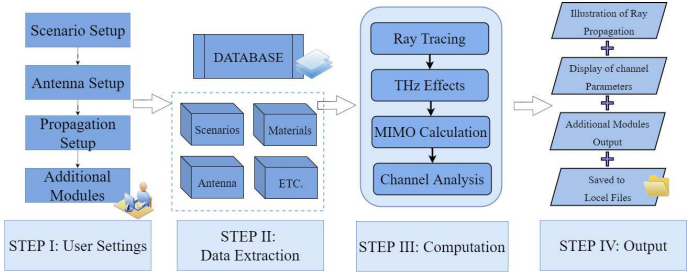}
	\caption{The workflow of a single simulation.} \label{2}
\end{figure}

\section{Simulation system setup}
\subsection{Model preparation for simulation}
Our simulation system supports 3 dimension (3-D) scenarios in the formats of \textit{json} (JavaScript Object Notation), \textit{stl} (stereolithography) and \textit{xml} (Extensible Markup Language). The office scenario for this paper is modeled with \textit{sketchup} 2021. The specific room dimensions are $7.2\times 7.2\times 3 m^3$. On one side of the wall it is equipped with two glass windows, which cover almost the entire wall on the other side. The rest of the walls are covered with wallpaper. The floor and celling is made of gypsum. The transmitting and receiving point are set at (1.46,2.42,2.41) and (5.2,5.2,1.5) respectively. The room is presented in Fig.~\ref{1}.

\subsection{Software architecture and workflow}
Constructed in MATLAB, our simulation system consists of five main modules, i.e., antenna, geometry, ray tracing, paths, and output module, which correspond to different functions. The workflow of system is vividly illustrated in Fig.~\ref{2}.

\subsection{Simulation acceleration based on parallel computation}
The intersection determination process in ray tracing in the second step in \ref{2.1.2} is extremely time-consuming, thus we employ parallel computation in our system for acceleration. Graphic processing unit (GPU) computation is especially suitable for RT acceleration with massive MIMO arrays \cite{9}. We implement kernel functions to achieve parallel computation, encapsulating compute unified device architecture (CUDA) C++ programs for GPU as a library to be called in MATLAB. With GPU acceleration, we can reduce the time consumption of massive MIMO simulations by above 81\%.
\section{Results and analysis}
\subsection{SBR ray tracing and calculation}

\begin{figure}[t]
	\centering
	\subfigure[]{
		\includegraphics[width=0.2\textwidth]{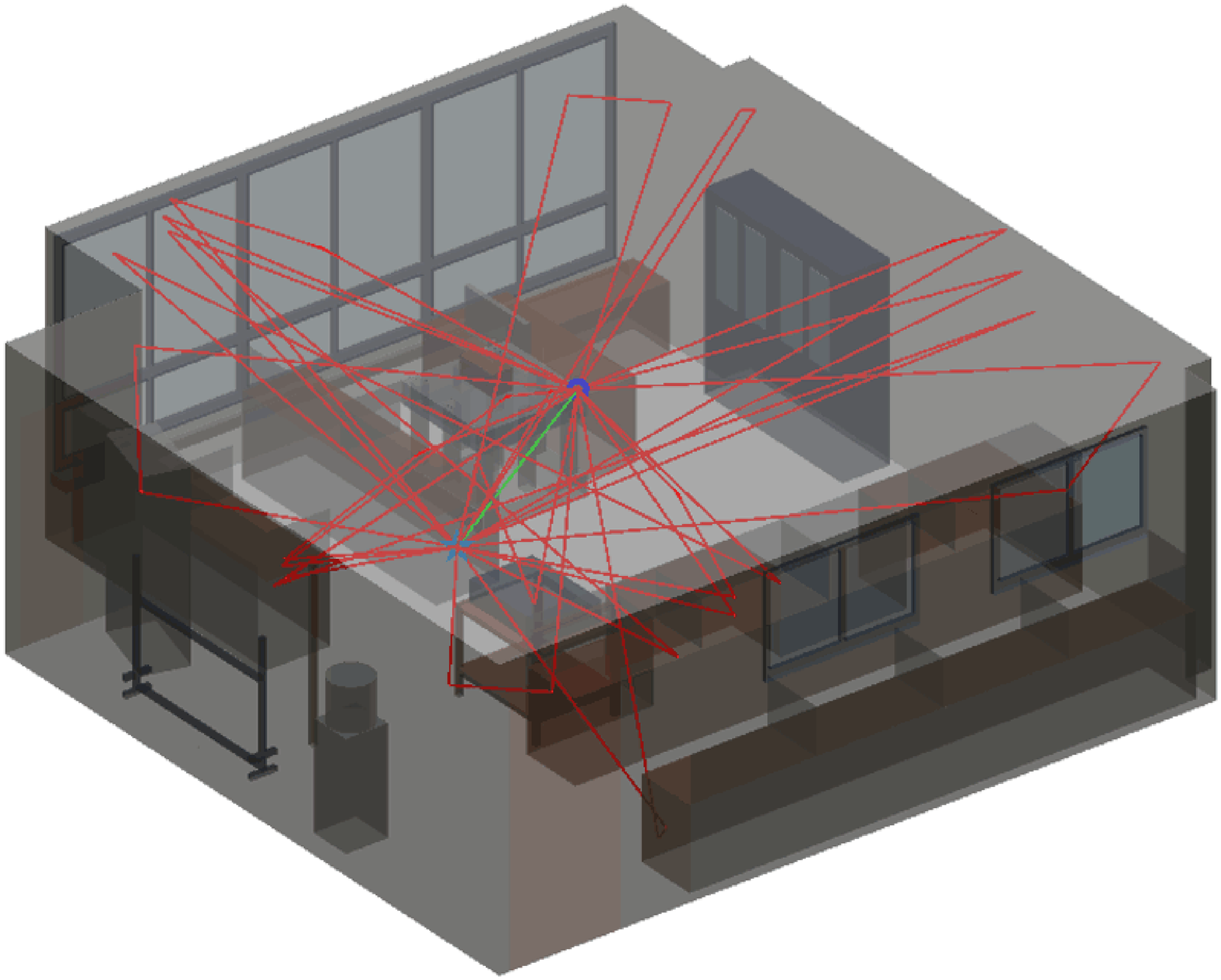}
	}
	\qquad
	\subfigure[]{
		\includegraphics[width=0.2\textwidth]{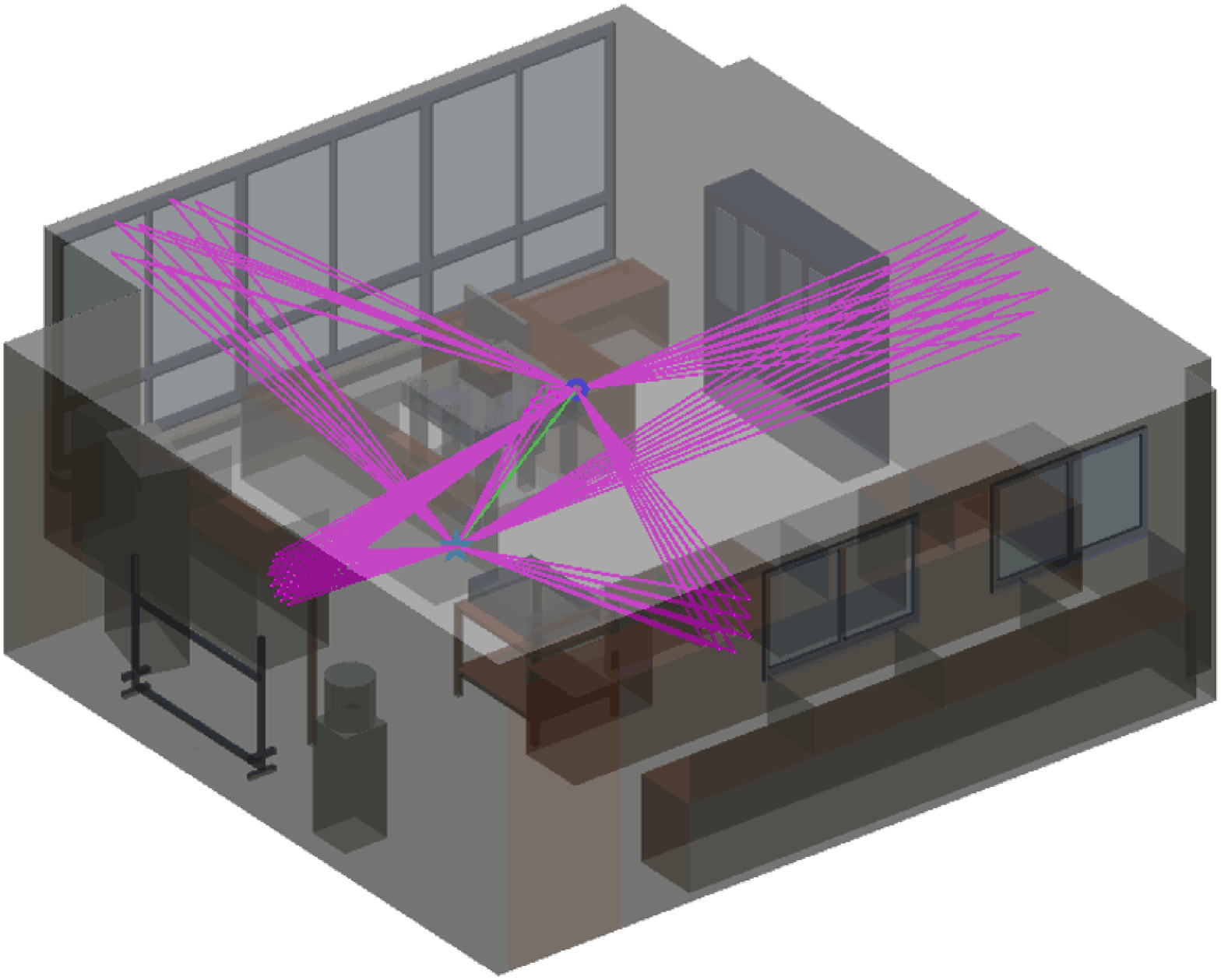}
	}
	\qquad
	\subfigure[]{
		\includegraphics[width=0.2\textwidth]{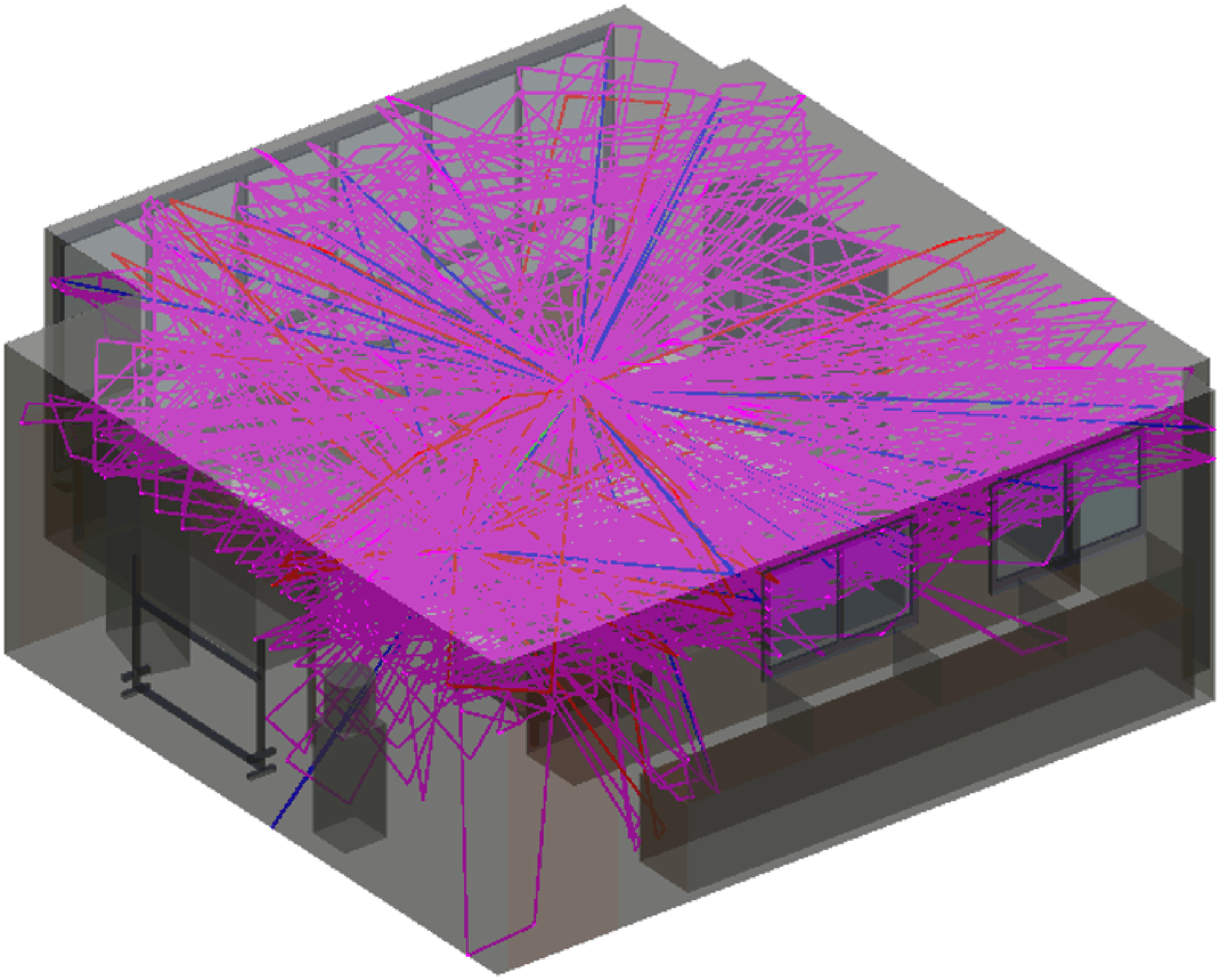}
	}
	\caption{The ray tracing results based on different settings, (a). Second order reflection, (b). Directive scattering model based on ER theory, and (c). First order diffraction.}
	\label{3}
\end{figure}

\begin{figure}[tb]
	\centering
	\subfigure[]{
		\includegraphics[width=0.35\textwidth]{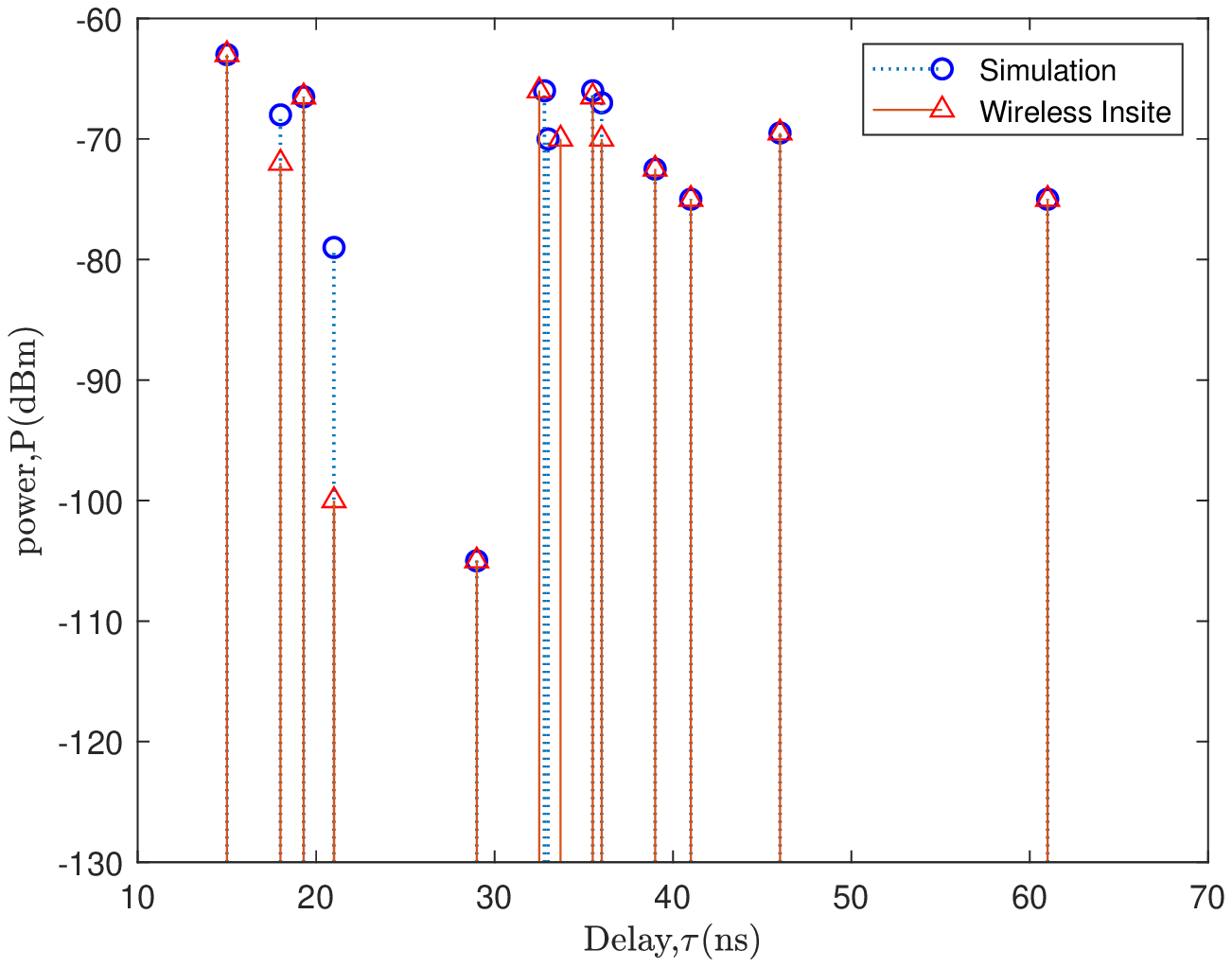}
	}
	\qquad
	\subfigure[]{
		\includegraphics[width=0.35\textwidth]{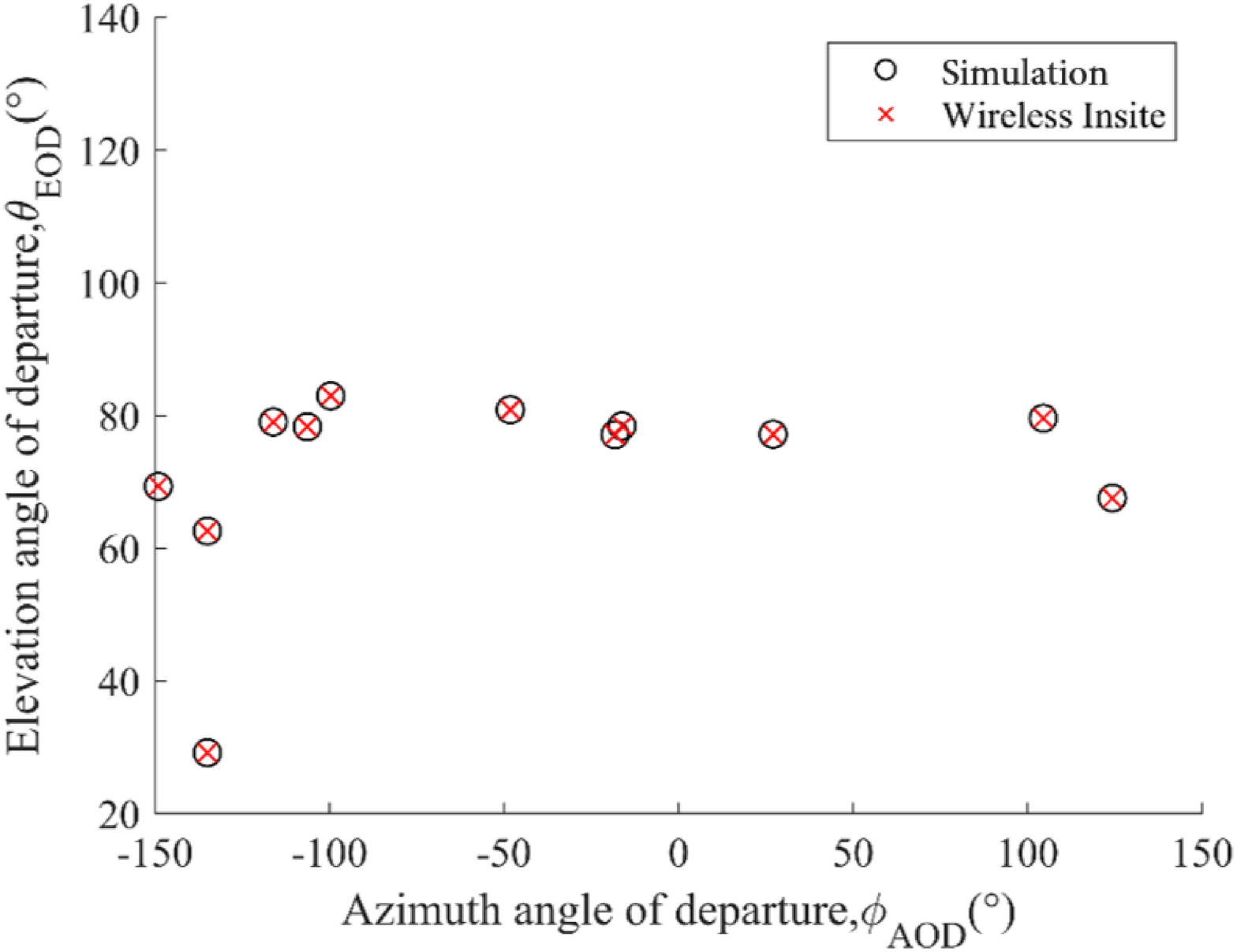}
	}
	\qquad
	\subfigure[]{
		\includegraphics[width=0.35\textwidth]{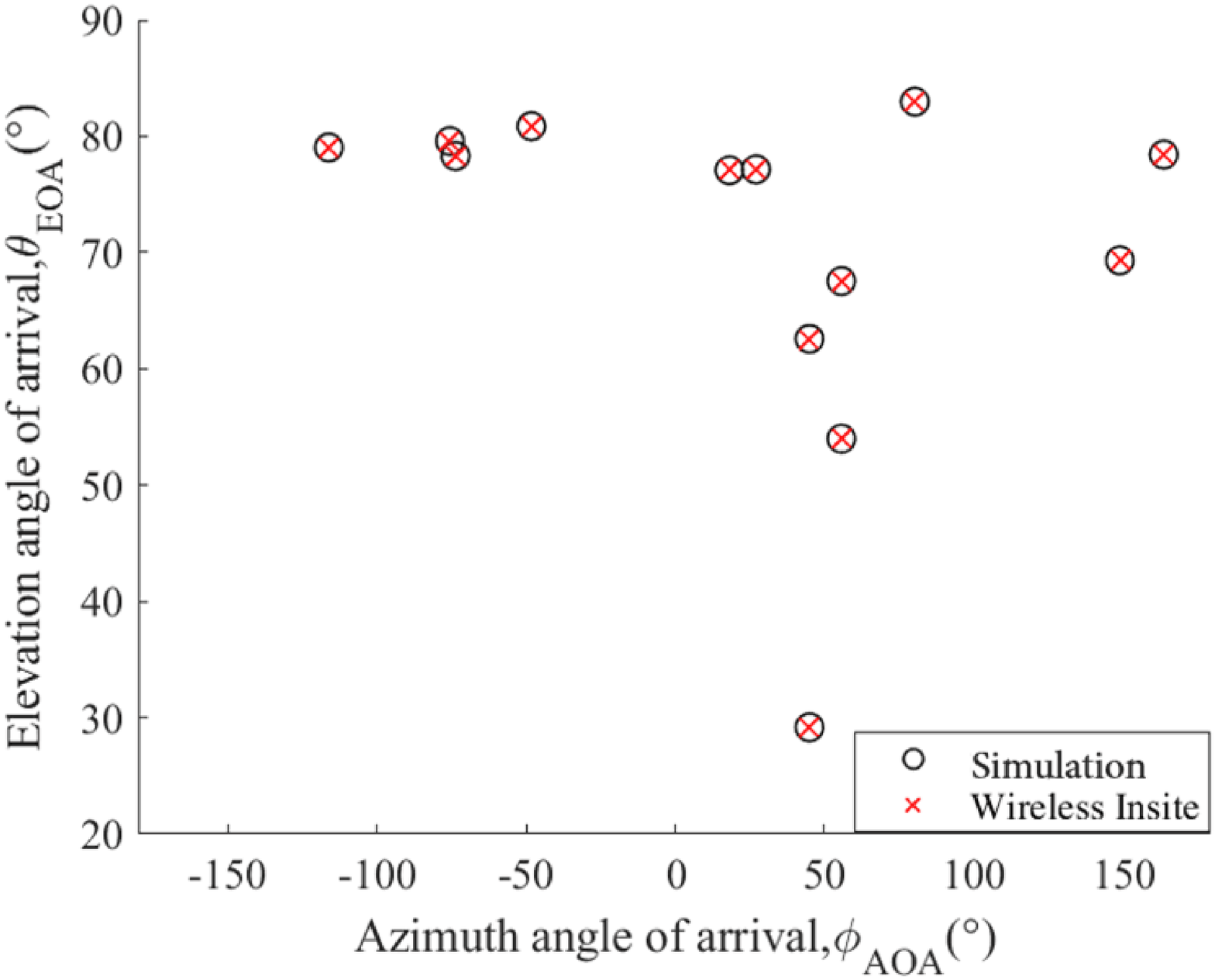}
	}
	\caption{Comparison between results simulated by this system and those by WI respectively, (a). PDP, (b). AoD, and (c). AoA.}

	\label{4}
\end{figure}

As is introduced above, our RT simulation supports line-of-sight (LoS) propagation as well as reflection, diffraction, and scattering. Some RT simulation results are displayed in Fig.~\ref{3} at 2.4 GHz with third order reflection. This frequency band is selected because it is supported by WI. Then the ray calculation results can be compared with those simulated with WI, and have reached fine agreements in Fig.~\ref{4} within 5.2\% error. The calculated results of channel characteristics, taking 2.4~GHz with forth order reflection and first order diffraction for example, are listed in Table~\ref{t1}.

\begin{table}[t]
	\centering
	\caption{The calculated simulation results}
	\label{t1}
	\begin{tabular}{|c|c|}
		\hline
		Parameter & Value \\
		\hline
		Pass loss (dB) & $83.18$ \\
		\hline
		Channel capacity (bit/(s $\times$ Hz)) & $52.516$ \\
		\hline
		RMS delay spread (ns) & $52.21$ \\
		\hline
		Mean angle of departure(AoD) $(\phi_t ,\theta_t)$ ($\degree$) & $(35.72,-16.90)$ \\
		\hline
		Mean angle of arrival(AoA) $(\phi_r ,\theta_r)$ ($\degree$) & $(37.91,15.83)$ \\
		\hline
	\end{tabular}
\end{table}

Such results are of great value for the design of actual communication systems. The path loss helps design channels with less energy loss, while the level of channel transmission rate can be estimated from the simulated channel capacity. RMS delay spread helps analyze whether the waveform expansion will cause inter-code crosstalk and evaluate its communication quality. The AoA and AoD can guide the beam assignment design of antenna, which is meaningful for the deployment of communication channel.

\begin{figure}[tbp]
	\centering
	\subfigure[]{
		\includegraphics[width=0.5\textwidth]{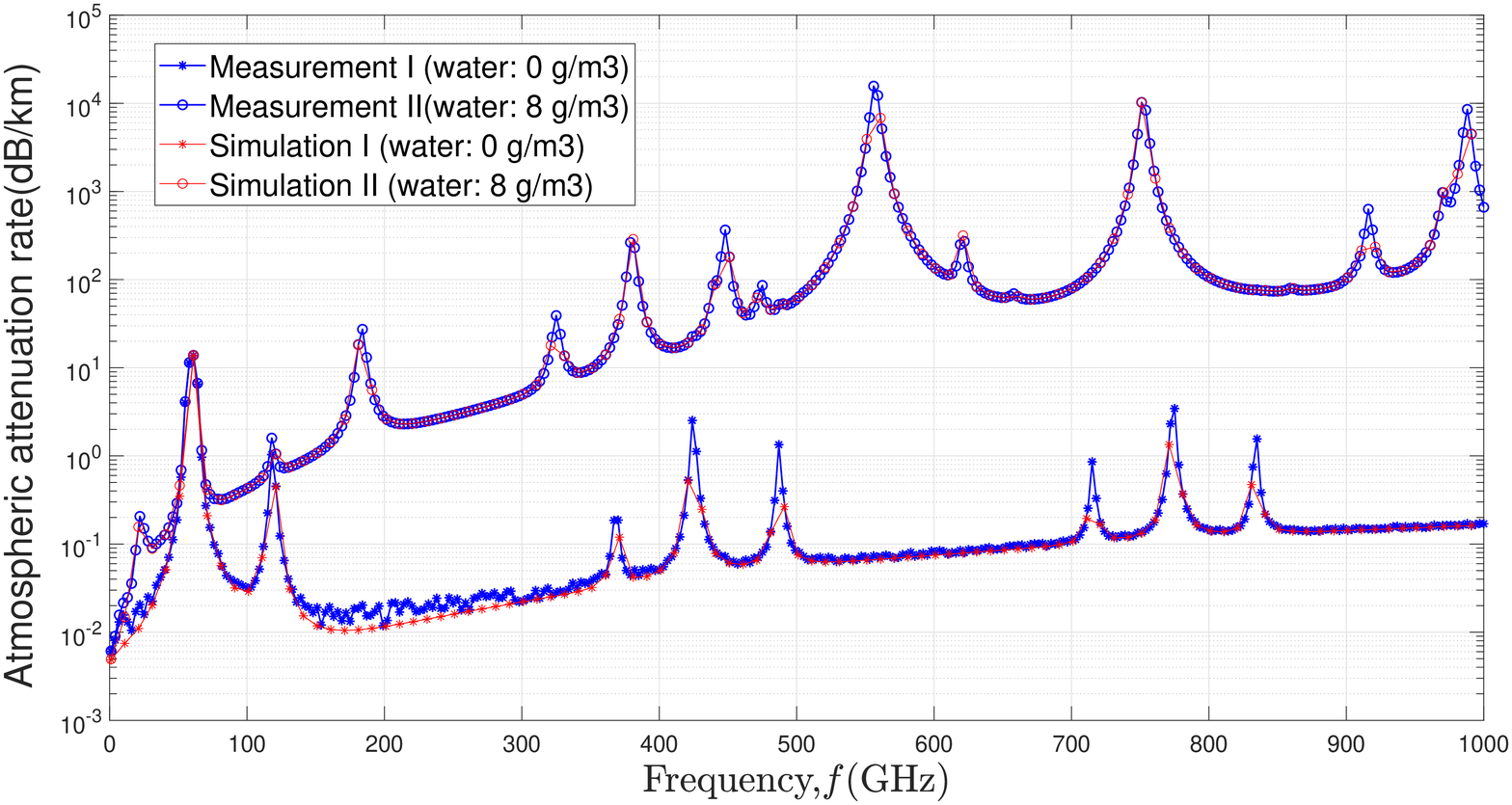}
	}
	\qquad
	\subfigure[]{
		\includegraphics[width=0.5\textwidth]{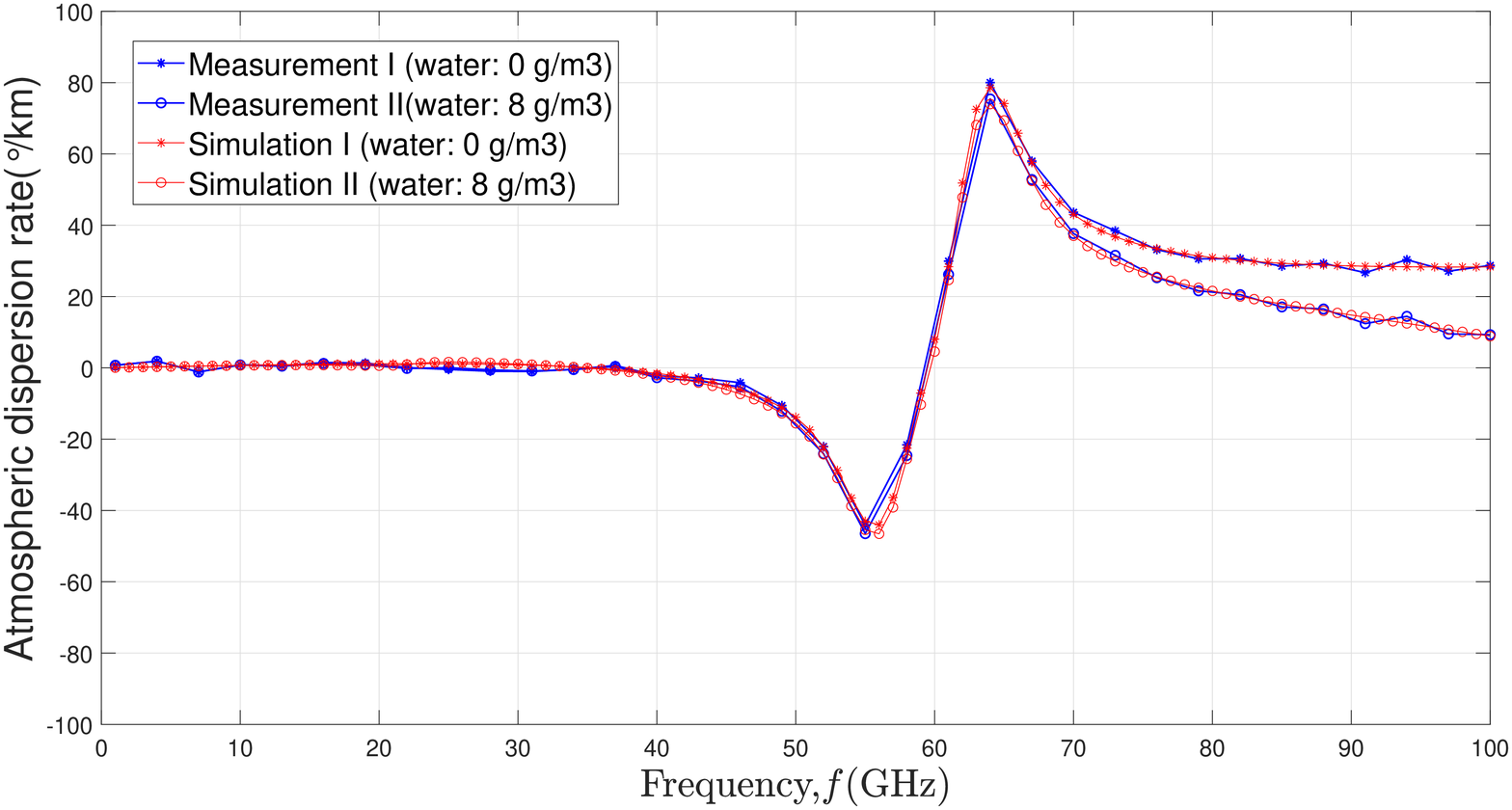}
	}
	\caption{The simulation results of atmospheric absorption phenomenon with the density of water vapor at the level of 0 g/$m^3$ and 8 g/$m^3$, and is
	compared with the results from ITU-R respectively, (a). Attenuation rate (dB/km), (b). Dispersion rate($\degree$/km).}
	\label{5}
\end{figure}
\subsection{THz band simulations}
\subsubsection{Air attenuation and dispersion simulations} The simulation for air attenuation and dispersion is demonstrated graphically in Fig.~\ref{5}, in which the results is in good consistency with the results
in ITU-R documents
\cite{6}. At higher frequencies, the molecular absorption of water vapor is dominant in atmospheric interference, which grows stronger with increasing humidity.
It is easy to figure out that each curve has absorption peaks at several specific signal frequencies, thus it is necessary to avoid these frequencies when actually deploying THz channels so that power wastage can be well mitigated.

\subsubsection{The Kirchhoff scattering based reflection modification} 
\begin{figure}[tb]
	\centering
	\includegraphics[width=0.5\textwidth]{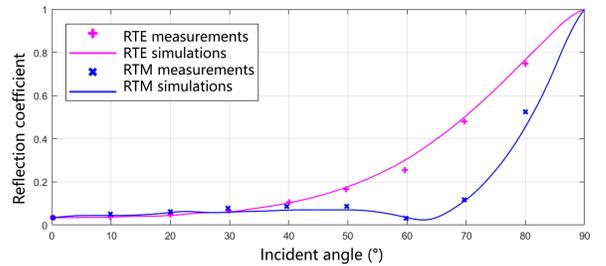}
	\caption{The simulation results of reflection modification compared with measurements.} \label{111}
\end{figure}

Comparison has been made between simulation and measurement for modified reflection coefficients of the wallpaper sample in the scenario. Good agreements are found in Fig.~\ref{111} with errors generally lower than 6.5\% in reflection coefficients simulations for TE and TM polarized waves at the frequency of 350 GHz.

\subsection{Massive MIMO simulations}
\begin{figure}[t]
	\centering
	\includegraphics[width=0.5\textwidth]{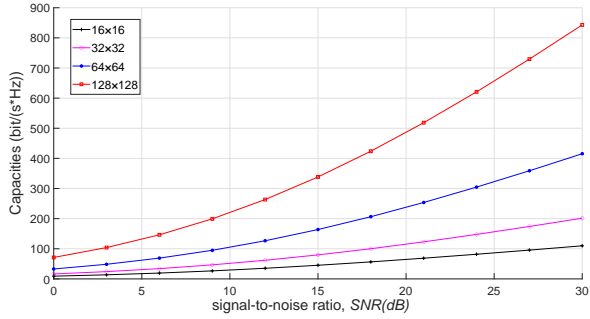}
	\caption{The simulated channel capacity for $16\times16$, $32\times32$, $64\times64$ and $128\times128$ MIMO.} \label{6}
\end{figure}

The simulation results of channel capacity for massive MIMO system are shown in Fig.~\ref{6}. This pattern indicates the changing trend of channel capacity, which increases rapidly as the scales of antenna arrays in MIMO systems grow. The result illustrates the improvement of MIMO channel capacity, demonstrating the inherent advantage of massive MIMO in terms of transmission rate improvement.

\begin{figure}[tbp]
	\centering
	\includegraphics[width=0.5\textwidth]{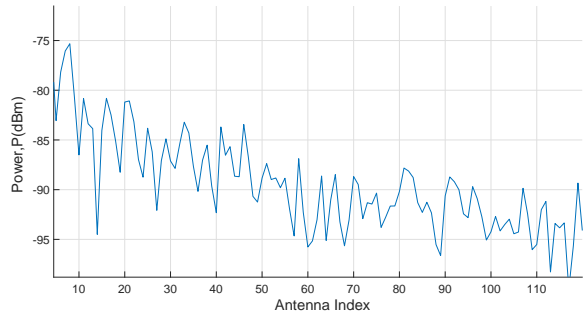}
	\caption{The received power of MIMO with the quantity of receivers varying from 1 to 128.} \label{7}
\end{figure}

The power under different antenna ports combined with phase is illustrated in Fig.~\ref{7}, where the power of received signals in MIMO array demonstrates spatial non-stationarity in distribution. This trend is in comply with the non-stationary characteristics of MIMO channels, which is caused by the shelter of LoS path between the receive point and one certain antenna. The shadow fading caused by determined scatterer in the scenario and the multipath effect collectively result in this pattern with a zig-zag like line.
\section{Conclusions}
In this paper, we have employed SBR method to model 6G THz and massive MIMO channels in a complex indoor scenario. Simulations were carried out with our proposed RT system and the results have been verified by commercial RT software WI within 5.2\% error. We have modified the calculation method of Fresnel reflection coefficient according to Kirchhoff scattering theory and the simulation results have been proved consistent with experimental data in literatures with an error lower than 6.5\%. The atmospheric absorption effect has also been considered to exemplify the frequency selective fading effect at THz band. MIMO simulation module has also been integrated in the system to analyze the capacity improvement and spatial non-stationarity brought by MIMO. Moreover, GPU parallel computation has been employed to accelerate MIMO simulations and the time consumption can be cut down to 19\%. The results of our work indicate that RT method is suitable for MIMO and THz channel model. In the future, more experimental data for materials under THz band and better spherical wave models for ultra-massive MIMO will be considered.
\section*{Acknowledgment}
\small {This work was supported by the National Key R\&D Program of China
under Grant 2018YFB1801101, the National Natural Science Foundation
of China (NSFC) under Grants 61960206006 and 61901109, the Frontiers
Science Center for Mobile Information Communication and Security, the High
Level Innovation and Entrepreneurial Research Team Program in Jiangsu, the
High Level Innovation and Entrepreneurial Talent Introduction Program in
Jiangsu, the Research Fund of National Mobile Communications Research
Laboratory, Southeast University, under Grant 2021B02, the High Level
Innovation and Entrepreneurial Doctor Introduction Program in Jiangsu under
Grant JSSCBS20210082, the Fundamental Research Funds for the Central
Universities under Grant 2242022R10067, and the EU H2020 RISE TESTBED2
project under Grant 872172.}
%\begin{figure}[t]
%	\centering
%	\includegraphics[width=0.5\textwidth]{Picture/fig7.eps}
%	\caption{The received power of MIMO with the quantity of receivers varying from 1 to 128.} \label{7}
%\end{figure}


\begin{thebibliography}{10}
	\bibitem{1}
	C. -X. Wang, J. Huang, H. Wang, X. Gao, X. You, and Y. Hao, ``6G wireless channel measurements and models: trends and challenges," \textit{IEEE Trans. Veh. Technol}, vol.~15, no. 4, pp.~22--32, Dec. 2020.
	\bibitem{2}
	X. -H. You, C. -X. Wang, J. Huang, et al., ``Towards 6G wireless communication networks: vision, enabling technologies, and new paradigm shifts," \textit{Sci. China Inf. Sci.}, vol.~64, no.1, Jan. 2021.
	\bibitem{b}
	A. Zhou, J. Huang, J. Sun, Q. Zhu, C. -X. Wang and Y. Yang, ``60 GHz channel measurements and ray tracing modeling in an indoor environment," \textit{Proc.  WCSP},  Nanjing, China, pp.~1--6, Oct. 2017.
	\bibitem{a}
	D. He, B. Ai, K. guan, L. Wang, Z. Zhong, and T. Kürner, ``The design and applications of high-performance ray-tracing simulation platform for 5G and beyond wireless communications: A tutorial," \textit{IEEE Commun. Surveys Tuts.}, vol.~21, no. 1, pp.~10--27, Firstquarter 2019.
	\bibitem{3}
	H. Ling, R. -C. Chou, and S. -W. Lee, ``Shooting and bouncing rays: calculating the RCS of an arbitrarily shaped cavity," \textit{IEEE Trans. Antennas Propag.}, vol.~37, no. 2, pp.~194--205, Feb. 1989.
	\bibitem{4}
	S. Kasdorf, B. Troksa, C. Key, J. Harmon, and B. M. Notaroš, ``Advancing accuracy of shooting and bouncing rays method for ray-tracing propagation modeling based on novel approaches to ray cone angle calculation," \textit{IEEE Trans. Antennas Propag.}, vol.~69, no. 8, pp.~4808--4815, Aug. 2021.
	\bibitem{5}
	C. Han, A. O. Bicen, and I. F. Akyildiz, ``Multi-ray channel modeling and wideband characterization for wireless communications in the terahertz band," \textit{IEEE Trans. wirel. commun.}, vol.~14, no. 5, pp.~2402--2412, May.2015.
	\bibitem{6}
	Rec. ITU-R P.676-12: Attenuation by atmospheric gases and related effects, ITU, 2019.
	\bibitem{7}
	G. Liu, J. She, W. Lu, et al., ``3D deterministic ray tracing method for massive MIMO channel modelling and parameters extraction," \textit{IET Commun.}, vol.~14, no.18, pp.~3169--3174, Oct. 2020.
	\bibitem{8}
	J. Wang, C. -X. Wang, J. Huang, et al., ``A novel 3D non-stationary GBSM for 6G THz ultra-massive MIMO wireless systems," \textit{IEEE Trans. Veh. Technol.}, vol.~70, no. 12, pp.~12312--12324, Dec. 2021.
	\bibitem{c}
	C. -X. Wang, Z. Lv, X. Gao, X. You, Y. Hao and H. Haas, ``Pervasive wireless channel modeling theory and applications to 6G GBSMs for all frequency bands and all scenarios", \textit{IEEE Trans. Veh. Technol}, pp.~1--1, Jun. 2022.
	\bibitem{9}
	J. Tan, Z. Su and Y. Long , ``A full 3-D gPU-based beam-tracing method for complex indoor environments propagation modeling,"\textit{IEEE Trans. Antennas Propag.}, vol.~63, no. 6, pp.~2705--2718, Jun. 2015.
\end{thebibliography}
\end{document}